\documentclass[10pt,a4paper]{article}

\usepackage[utf8]{inputenc}
 \usepackage{charter}
\usepackage{amsfonts}
\usepackage{amssymb}
\usepackage{amsmath}
\usepackage{amsthm}
\usepackage{mathrsfs}
\usepackage{verbatim}
\usepackage{graphicx}
\usepackage{epstopdf}
\usepackage{enumitem}
\usepackage{caption}
\usepackage{subcaption}
\usepackage{appendix}
\usepackage{xcolor}
\usepackage{bm}

\theoremstyle{remark} \newtheorem{remark}{Remark}

\theoremstyle{remark}

\newcommand{\Xvec}{\mathbf{X}}
\newcommand{\Yvec}{\mathbf{Y}}

\newcommand{\realSet}{\mathcal{R}}

\newcommand{\E}{\mathbb{E}}

\newcommand{\argmin}{\operatornamewithlimits{argmin}}

\renewcommand{\vec}[1]{{\mathbf{#1}}}

\begin{document}

\title{$k$-NN Estimation of Directed Information}

\author{Yonathan Murin\\ 
 {\small Department of Electrical Engineering, Stanford University, USA} \\
 {\small {\em moriny@stanford.edu}}}
\date{}


\maketitle

\begin{abstract}
	This report studies data-driven estimation of the directed information (DI) measure between two{em  discrete-time and continuous-amplitude} random process, based on the $k$-nearest-neighbors ($k$-NN) estimation framework.
	Detailed derivations of two $k$-NN estimators are provided. 
	The two estimators differ in the metric based on which the nearest-neighbors are found. 
	To facilitate the estimation of the DI measure, it is assumed that the observed sequences are (jointly) Markovian of order $m$. As $m$ is generally not known, a data-driven method (that is also based on the $k$-NN principle) for estimating $m$ from the observed sequences is presented. 
	An exhaustive numerical study shows that the discussed $k$-NN estimators perform well even for relatively small number of samples (few thousands). Moreover, it is shown that the discussed estimators are capable of accurately detecting linear as well as non-linear causal interactions.
\end{abstract}

\section{Introduction}	

Detection and estimation of causality relationships between two random processes is a fundamental problem in many natural and social sciences \cite{Hlayackova07}. 
This task is in particular challenging as in many real-life scenarios one does not have a good underlying statistical model for the considered process, e.g., in the fields of neuroscience, financial markets, meteorology, etc. For such scenarios it is desirable to use a non-parametric estimator for the {\em causal influence} between two observed time-series.
The common approach for quantifying the causal influence between two time-series dates back to the seminal work of Granger \cite{Granger69} where $\{ X_n \}$ is said to have a Granger-causal influence on $\{ Y_n \}$ if:

\smallskip
 {\slshape Given the past of $Y_n$, the past of $X_n$ helps in predicting future samples of $Y_n$, e.g., $Y_{n+1}$.} 

\smallskip
\noindent While this approach is indeed general, the common formulation of Granger causality (GC) assumes that the time-series obey a linear structure, and therefore it does not follow the non-parametric approach mentioned above. 
A possible alternative to GC is the information theoretic measure of directed information (DI) \cite{massey90, kramerThesis}, which is closely
related to the transfer entropy (TE) functional \cite{wibral2014}.

In this report we discuss the estimation of DI between two discrete-time and {\em continuous-amplitude} sequences (time-series). The estimated DI can then be used as a measure of the causal influence between the random processes underlying the observed sequences.\footnote{Note that the work \cite{Janzig13} presented scenarios where DI (or TE) fail to detect or quantify the causal influence between a pair of time-series. The conclusions of \cite{Janzig13} also hold for the GC measure.} 
Note that DI is a deterministic function of the joint density of the underlying random processes.
Therefore, the DI can be estimated by first estimating the (local) joint densities, and then using these densities to estimate the DI. This approach was taken in \cite{Malladi2016} where it was proposed to use a kernel density estimator (KDE) for estimating the local densities, and in \cite{Sabesan2009} that suggested to estimate the local the densities using correlation integrals. On the other hand, the estimation method we discuss in the current report is based on the $k$-nearest-neighbors ($k$-NN) principle. 
Before delving into the technical details of estimating the DI functional, we emphasize that when estimating statistical functionals it is common to assume that the underlying process are {\em stationary, ergodic, and smooth}. In the rest of this report we build upon these assumptions without verifying their validity. 

The rest of this report is organized as follows: In Section \ref{sec:DiffEnt} we discuss $k$-NN estimation of differential entropy. This estimation approach is extended to estimating the mutual information (MI) functional in Section \ref{sec:MutInf}. Estimating the DI is discussed in Section \ref{sec:DirectInf}, and a numerical study is presented in Section \ref{sec:numerical}.

{\bf {\slshape Notation}:} We denote random variables (RVs) by upper case letters, $X$, and their realizations with the corresponding lower case letters. We use the short-hand notation $X_1^n$ to denote the sequence $\{X_1,X_2,\dots,X_n\}$. We denote random processes using boldface letters, e.g., $\mathbf{X}$. We denote sets by calligraphic letters, e.g., $\mathcal{S}$, where $\realSet$ denotes the set of real numbers. $f_X(x)$ denotes the probability density function (PDF) of a continuous RV $X$ on $\realSet$, and $\log(\cdot)$ denotes the natural basis logarithm. Finally, we use $h(\cdot)$ and $I(\cdot;\cdot)$ to denote differential entropy and mutual information as defined in \cite[Ch. 8]{cover-book}.

\section{$k$-NN Estimation of Differential Entropy} \label{sec:DiffEnt}

We introduce the concept of $k$-NN estimation of information theoretic measures by first discussing the estimation of differential entropy. A detailed discussion regarding methods (including $k$-NN) for estimating differential entropy, MI, and the Kullback–Leibler (KL) divergence is provided in \cite{Wang08}. 
A popular approach for estimating information theoretic functionals is the {\em re-substitution} method where first the density is {\em locally} (around each of the data points) estimated, and then the functional is estimated via empirical averaging \cite[Sec. 2.2.1]{Wang08}. 

Specifically, let $X_i \in \realSet^d, i=1,2,\dots,N$, be independent and identically distributed (i.i.d.) samples of $X$ with PDF $f_X(x)$, and consider estimating the differential entropy of $X$ from $X_1^N$. 
Let $\hat{f}_X(X_i), i \in \{1,2,\dots,N\}$, be a {\em local} estimation of the density around the $i^{\text{th}}$ sample. Recalling that the differential entropy is defined as $h(X) \triangleq - \E \{ \log f_{X}(x)\}$, see \cite[Ch. 8]{cover-book}, the re-substitution estimator $\hat{h}(X)$ of $h(X)$ is given by:
\begin{align}
	\hat{h}(X) = - \frac{1}{N} \sum_{i=1}^N \log \hat{f}_{X}(X_i). \label{eq:resubEntropy}
\end{align}

While there are several popular techniques for estimating the local density, e.g., kernel density estimation \cite{Moon95} and correlation integrals \cite{Grassberger83}, in this report we focus on estimation algorithms based on the $k$-NN principle. 
Before presenting the estimation strategy, we provide several definitions. Let $p\ge 1$. For $x \in \realSet^d$, the $\ell_p$-norm of $x$ is defined as:
\begin{align*}
	|| x ||_p \triangleq \left( \sum_{i=1}^d |x_i|^p \right)^{\frac{1}{p}},
\end{align*}

\noindent where the $\ell_p$-distance between $x_i \in \realSet^d$ and $x_j \in \realSet^d$ is defined as $d_p(x_i, x_j) \triangleq || x_i - x_j ||_p$. 
Let $\mathsf{sort}(\vec{x}), \vec{x} \in \realSet^{n}$, denote the sorted version of the vector $\vec{x}$, in ascending order, and define $\bm{\rho}_{i,p} \triangleq \mathsf{sort}\left( \{ d_p(x_i, x_j) \}_{j=1, j\neq i}^N  \right)$ to be the (sorted) vector of distances, in $\ell_p$-norm, of all the samples from $x_i$. In particular, $\rho_{k,i,p}$ denotes the distance from $x_i$ to its $k$-NN.\footnote{Unless otherwise stated, in this report we assume that $k$ is a fixed and relatively small number (independent of $N$) at the range $4,5,\dots,10$.} 

Assuming a {\em uniform local} density in a small environment around each of the samples (this assumption implies that the density $f_X(x)$ is smooth), $\rho_{k,i,p}$ can be used to estimate $f_X(X_i)$. Let $\Gamma(a)$ denote the Euler's gamma function \cite[eq. (5.2.1)]{nist10}, and let $c_{d,p}$ denote the volume of the unit $l_p$-ball in $d$ dimensions, given by \cite{Wang05}: 
\begin{align*}
	c_{d,p} = 2^d \frac{(\Gamma(1 + \frac{1}{p}))^d}{\Gamma(1 + \frac{1}{p})}.
\end{align*}

\noindent Since in the ball of radius $\rho_{k,i,p}$ there are $k$ samples out of $N$ in total, we can approximate the local density via:
\begin{align}
	\hat{f}_X(X_i) \approx \frac{k}{N c_{d,p} (\rho_{k,i,p})^d}. \label{eq:localDensity}
\end{align}

\noindent Substituting \eqref{eq:localDensity} into \eqref{eq:resubEntropy} we obtain the following entropy estimator:
\begin{align}
	\hat{h}(X) = \frac{1}{N} \sum_{i=1}^{N} \log \left( \frac{N c_{d,p} (\rho_{k,i,p})^d}{k} \right). \label{eq:EntropyEstBasic}
\end{align}

\noindent The work \cite{KL87}, by Kozachenko and Leonenko (KL), showed that the simple estimator \eqref{eq:EntropyEstBasic} is biased, and suggested the following bias-corrected estimator:
\begin{align}
	\hat{h}_{\text{KL}}(X) & = \frac{1}{N} \sum_{i=1}^{N} \left\{ \log \left( \frac{N c_{d,p} (\rho_{k,i,p})^d}{k} \right) + \log(k) - \psi(k) \right\} \nonumber \\
	& = \log(N) - \psi(k) + \log(c_{d,p}) + \frac{d}{N} \sum_{i=1}^{N} \log(\rho_{k,i,p}), \label{eq:KLEst}
\end{align}

\noindent where $\psi(k)$ is the digamma function \cite[Ch. 5.4]{nist10}.

\begin{remark}[{\em The case of $k$ dependent of $N$}]
	If $k$ is chosen as a function of $N$, then $\psi(k)$ converges to $\log(k)$ and no bias-correction is required to obtain a consistent estimator. On the other hand, if $k$ is fixed, then the correction term $\log(k) - \psi(k)$ is crucial for consistency.
\end{remark}

Next, we discuss $k$-NN estimation of mutual information (MI).

\section{From Entropy to Mutual Information} \label{sec:MutInf}

Consider two RVs $X \in \realSet^{d_x}$ and $Y \in \realSet^{d_y}$. Observing $N$ i.i.d pairs $(X_i,Y_i)$ from the joint density $f_{X,Y}(x,y)$, we are interested in estimating the MI $I(X;Y) = h(X) + h(Y) - h(X,Y)$. 
Fixing $k$ and estimating each of the entropy terms via $\hat{h}_{\text{KL}}(\cdot)$, one obtains the following consistent estimator (the consistency follows from the consistency of the $\hat{h}_{\text{KL}}(\cdot)$ estimator), denoted by $\hat{I}_{3\text{KL}}(X;Y)$:
\begin{align}
	\hat{I}_{3\text{KL}}(X;Y) = \hat{h}_{\text{KL}}(X) + \hat{h}_{\text{KL}}(Y) - \hat{h}_{\text{KL}}(X,Y). \label{eq:MI3KL}
\end{align}   

\noindent Note that in the estimator \eqref{eq:MI3KL} each entropy term is estimated {\em separately}, with a {\em different} $k$-NN distance. Thus, even though the estimator \eqref{eq:MI3KL} is consistent, since the estimators are not coupled, for a {\em finite number of samples} the bias can be non-negligible. This motivated the work of Kraskov, St\"{o}gbauer and Grassberger (KSG) \cite{KSG}, that presented a modification of the $\hat{I}_{3\text{KL}}(X;Y)$ estimator, and {\em empirically} showed that this modification improves performance (higher accuracy) when the number of samples is finite. In the following we refer to this estimator as $\hat{I}_{\text{KSG}}(X;Y)$.

The main idea behind $\hat{I}_{\text{KSG}}(X;Y)$ is to modify the estimators of the individual entropy terms, $h(X)$ and $h(Y)$, such that their correlation with the estimator of the joint entropy $h(X,Y)$ term will be higher, leading to a smaller bias.
This interpretation was recently presented in \cite{GaoDemens16}.
Let $\rho_{k,i,p}(X,Y)$ denote the distance from {\em the pair} $(X_i, Y_i)$ to its $k$-NN. Further define $\mathbb{I}(\cdot)$ to be the indicator function and let $n_{x,i,p} \triangleq \sum_{j=1,j\neq i}^N \mathbb{I}(d_p(x_i, x_j) \le \rho_{k,i,p}(X,Y))$. $n_{y,i,p}$ is defined similarly. 
Note that $n_{x,i,p}$ is the number of samples which are within a distance $\rho_{k,i,p}(X,Y)$ from $X_i$, where the {\em distance is measured in the $X$-plane}. Thus, as this places no limitation on the distance in the $Y$-plane, it follows that $n_{x,i,p} \ge k$. Using these definitions, and setting $p = \infty$, the KSG estimator is given by:
\begin{align}
	\hat{I}_{\text{KSG}}(X;Y) = \psi(k) + \log(N) - \frac{1}{N} \sum_{i=1}^N (\psi(n_{x,i,\infty} + 1) + \psi(n_{y,i,\infty} + 1)). \label{eq:MIKSG}
\end{align}

\noindent Note that when $p = \infty$, then $d_{X,Y} = d_X + d_Y, c_{d_{X,Y},\infty} = c_{d_{X},\infty} \cdot c_{d_{Y},\infty}$, and $\rho_{k,i,\infty}(X,Y) = \max \{\rho_{k,i,\infty}(X) , \rho_{k,i,\infty}(Y) \}$. Hence, the KSG estimator in \eqref{eq:MIKSG} estimates the {\em individual} entropy $h(X)$ via:
\begin{align}
	\hat{h}_{\text{KSG}}(X) = \log(N) + \log(c_{d_X,\infty}) + \frac{1}{N} \sum_{i=1}^{N} \left( d_X \log(\rho_{k,i,\infty}(X,Y)) - \psi(n_{x,i,\infty} + 1) \right), \label{eq:EntKSG}
\end{align}

\noindent which is {\em sample dependent}. On the other hand, the estimation of the joint entropy term is identical to the one in \eqref{eq:KLEst}. The work \cite{GaoDemens16} showed that $\hat{I}_{\text{KSG}}(X;Y)$ is consistent and derived the order of its bias. 

\begin{remark}[{\em The inherent bias in \eqref{eq:EntKSG}}] \label{rem:inherentBias}
	The estimator \eqref{eq:MIKSG} uses a {\em hyper-cube} of radius $\rho_{k,i,p}(X,Y)$ around each sample point. Since $\rho_{k,i,\infty}(X,Y) = \max \{\rho_{k,i,\infty}(X) , \rho_{k,i,\infty}(Y) \}$, it follows that for one of the individual terms, e.g., $X$, $\rho_{k,i,\infty}(X,Y)$ is {\em exactly} the distance to the $(n_{x,i,\infty} + 1)^{\text{th}}$-NN, while for the other term this is not the case, see \cite[Fig. 1]{KSG} for illustration of this observation. Thus, letting $(X_i,Y_i)_k$ be the $k-$NN of $(X_i,Y_i)$, and assuming that $(X_i,Y_i)_k$ lies on the $X$-boundary of the hyper-cube around $(X_i,Y_i)$, the bias of the estimator \eqref{eq:EntKSG} is of the order of $\frac{1}{n_{y,i,\infty}}$.
	This inherent bias is partially addressed in the second estimator introduced in \cite[eqn. (9)]{KSG}. This estimator uses a {\em hyper-rectangle} instead of a hyper-cube, yet, using hyper-rectangles requires other approximations. Thus, none of the two estimators presented in \cite{KSG} is uniformly better than the other. We refer the reader to \cite[Fig. 3]{KSG} for a detailed discussion regarding the approximations used as apart of the second estimator of \cite{KSG}.
\end{remark}

Motivated by the idea of using the {\em sample dependent} $n_{x,i,\infty}$ and $n_{y,i,\infty}$, the work \cite{GaoDemens16} proposed a different method to tackle the inherent bias discussed in Remark \ref{rem:inherentBias}. The idea is to use an $\ell_2$ ball instead of the $\ell_{\infty}$ ball used in \eqref{eq:MIKSG}. With this choice $\rho_{k,i,2}(X,Y)$ is not the $X$-boundary distance nor the $Y$-boundary distance, see \cite[Fig. 5]{GaoDemens16}. 
However, when the Euclidean norm is used, namely $p=2$, a new relationship should be derived between $n_{x,i,2}$ (number of points in the $X$-space) and $f_X(x)$. Such a relationship is formulated in \cite[Thm. 9]{GaoDemens16}, which essentially states the following approximation:
\begin{align}
	\E \{ n_{x,i,2} \} \approx N f_X(x) c_{d_X, 2} (\rho_{k,i,2}(X,Y))^{d_X}.
\end{align}

\noindent This approximation motivates estimating $\log f_X(x)$ via:
\begin{align}
	\widehat{\log f_X(x)} = \log (n_{x,i,2}) - \log N - \log  c_{d_X, 2} - d_X \log \rho_{k,i,2}(X,Y).
\end{align}

\noindent Plugging this estimation to the re-substitution \eqref{eq:resubEntropy} results in the Gao-Oh-Viswanath (GOV) entropy estimator:
\begin{align}
	\hat{h}_{\text{GOV}}(X) = \log(N) + \log(c_{d_X,2}) + \frac{1}{N} \sum_{i=1}^{N} \left( d_X \log(\rho_{k,i,2}(X,Y)) - \log(n_{x,i,2}) \right). \label{eq:EntGOV}	
\end{align}

\noindent Finally, to obtain an estimator for the MI, we recall that $d_{X,Y} = d_X + d_Y$ and $c_{d_{X,Y},2} = c_{d_{X} + d_{Y},2}$. Using the estimator \eqref{eq:KLEst} to estimate the joint entropy and the estimators \eqref{eq:EntGOV} for the individual entropy terms one obtains:
\begin{align}
	\hat{I}_{\text{GOV}}(X;Y) 
		& = \log(N) + \psi(k) + \log \left( \frac{c_{d_X,2} \cdot c_{d_Y,2}}{c_{d_X + d_Y,2}} \right) - \frac{1}{N} \sum_{i=1}^{N} ( \log(n_{x,i,2}) + \log(n_{y,i,2})). \label{eq:MIGOV}
\end{align}

\noindent Comparing \eqref{eq:MIGOV} and \eqref{eq:MIKSG}, $\log \left( \frac{c_{d_X,2} \cdot c_{d_Y,2}}{c_{d_X + d_Y,2}} \right)$ can be viewed as a correction term for using the Euclidean norm. Finally, we note that the consistency of $\hat{I}_{\text{GOV}}(X;Y)$ is stated in \cite[Thm. 10]{GaoDemens16}.

\section{Directed Information} \label{sec:DirectInf}

\subsection{Definitions and Background}

Let $\Xvec$ and $\Yvec$ be arbitrary discrete-time continuous-amplitude random processes, and let $X^N \in \realSet^N$ and $Y^N \in \realSet^N$, be $N$-length sequences. The {\em directed information} from $X^N$ to $Y^N$ is defined as \cite{massey90}:
\begin{align}
	I(X^N \to Y^N) & \triangleq \sum_{i=1}^n I(X_1^i; Y_i | Y_1^{i-1}) \nonumber \\
	& = \sum_{i=1}^N h(Y_i | Y_1^{i-1}) - h(Y_i | Y_1^{i-1}, X_1^i) \nonumber \\
	& \stackrel{(a)}{=} h(Y^N) - h(Y^N || X^N), \label{eq:DIdef}
\end{align}

\noindent where (a) follows by defining $h(Y^N || X^N) \triangleq \sum_{i=1}^N h(Y_i | Y_1^{i-1}, X_1^i)$, and $h(\cdot)$ is the differential entropy. This definition implies that $I(X^N \to Y^N) = 0$ when $Y_i$ is independent of $X_1^i$, given $Y_1^{i-1}$.
DI can be viewed as quantifying the {\em causal influence} of the sequence $X^N$ on the sequence $Y^N$. Therefore, it is not surprising that in contrast to MI, DI is {\em not} symmetric. 
The directed information {\em rate} \cite{kramerThesis} between the processes $\Xvec$ and $\Yvec$ is defined as:
\begin{align}
	I(\Xvec \to \Yvec) \triangleq \lim_{N \to \infty} \frac{1}{N} I(X^N \to Y^N), \label{eq:DIrate}
\end{align}

\noindent provided that this limit exists. We now make the following assumptions regarding the processes $\Xvec$ and $\Yvec$.

%
%
%

\begin{enumerate}[label = {\bf A{\arabic*}})]
	\item \label{assmp:stationarity}
		The random processes $\Xvec$ and $\Yvec$ are assumed to be stationary, ergodic, and Markovian of order $m$ in the observed sequences. The {\em stationarity} assumption implies that the statistics of the considered random processes is constant throughout the observed sequences. 
		Note that formally speaking, the random processes $\Xvec$ and $\Yvec$ should be stationary in order to ensure the existence of the DI rate. From practical perspective, it is required that the sequences $X^N$ and $Y^N$ will be stationary to ensure that the causal influence does not change over the observed sequences. 
		{\em Ergodicity} is assumed to ensure that the observed sequences truly represent the underlying processes. 
		Finally, the {\em Markovity} assumption is common in modeling {\em real-life systems which have finite memory}. For example, \cite{joseph12, wissel13, Malladi2016} used Markov models in analysis of neural recordings, \cite{Erlwein_thesis} and \cite{dias09} used Markov models in financial modeling, and \cite{Snijders01}--\cite{Heaukulani2013} in social networks dynamics.
		We formulate the assumption of Markovity of order $m$ in the observed sequences via:
		\begin{align}
			\forall i > m,  f(y_i | Y_{1}^{i-1}) = f(y_i | Y_{i-m}^{i-1}) \text{ and } f(y_i | Y_{1}^{i-1}, X_{1}^{i}) = f(y_i | Y_{i-m}^{i-1},X_{i-m}^{i-1}). \label{eq:MarkovDef}
		\end{align}
		
	\noindent In \eqref{eq:MarkovDef} we use the simplifying assumption that the dependence of $y_i$ on past samples of $Y_{1}^{i-1}$ and $X_{1}^i$ is of the same order. These definitions and the estimation methods defined in the sequel can be easily extended to two different orders. Moreover, in \eqref{eq:MarkovDef} we implicitly assume that given $(Y_{i-m}^{i-1},X_{i-m}^{i-1})$, $y_i$ is independent of $X_i$ which reflects a setting where $X_i$ and $Y_i$ are simultaneously measured.
			
	\item \label{assmp:bounded_1}
		The entropy of the first sample $y_1$ exists, i.e., $|h(Y_1)| < \infty$.
	
	\item	\label{assmp:bounded_2}
		The following conditional entropy exists: $|h(Y_{m+1}|Y_{1}^{m}, X_{1}^{m})| < \infty$.
	
\end{enumerate} 

Assumptions \ref{assmp:bounded_1} and \ref{assmp:bounded_2} are required to mathematically insure that the DI rate exists and is equal to a simple expression that depends on the finite memory length $m$. Moreover, Assumptions \ref{assmp:bounded_1} and \ref{assmp:bounded_2} prevent the degenerate case of deterministic $Y_1$ or deterministic relationship between $Y_{m+1}$ and $Y_{1}^{m}, X_{1}^{m}$ (this is one of the scenarios discussed in \cite{Janzig13}). 
\noindent Under these assumptions, \cite[Lemmas 3.1 and 3.2]{Malladi2016} imply that $I(\Xvec \to \Yvec)$ exists and is equal to:
\begin{align}
	I(\Xvec \to \Yvec) 
	& = I(X_{i-m}^{i-1}; Y_{i}|Y_{i-m}^{i-1}), \quad i > m, \label{eq:DIrateContMarkov}
\end{align}

\noindent 
In view of \eqref{eq:DIrateContMarkov}, $I(\Xvec \to \Yvec)$ can be given the following interpretation: 

\smallskip
{\slshape Given the past of the sequence $Y$, namely $Y_1^{i-1}$, how much the past of the sequence $X$, namely $X_1^{i-1}$, helps in predicting the next sample of $Y$, $Y_i$?} 

It can be observed that \eqref{eq:DIrateContMarkov} is a function of $m$, the Markov order. 
When one is interested in a data-driven estimator of the DI, $m$ is unknown and {\em must} be estimated from the observed sequences. 

\subsection{Estimating the Markov Order $m$} \label{subsec:MarkovOrdEst}

While it is a reasonable assumption that the observed time-series have finite memory and thus obey a Markov model, the order $m$ should be estimated from the data. 

A possible approach for estimating $m$ is to choose the value that facilitates the {\em best prediction} of future samples (see \cite{Murin17BME} and references therein). 
Specifically, let $\mathcal{M}$ be a (finite) set of candidate Markov orders. $\hat{m}$ is estimated to be the value in this set that minimizes a pre-defined loss function in predicting the next sample of $Y_{1}^N$ from $\hat{m}$ previous samples. 
While in \cite{wibral2014} it was proposed to use the prediction method of \cite{Ragwitz2002}, i.e., use $k$-NN prediction of the next sample in $Y_{1}^N$ based on the past samples of $Y_{1}^N$, {\em this approach ignores the dependency} between $Y_{1}^N$ and $X_{1}^N$.
To account for this dependency, we propose to predict the next sample of $Y_{1}^N$ based on the past $\hat{m}$ samples of both $Y_{1}^N$ and $X_{1}^N$. 

Let $\varphi: \realSet^{2M} \to \realSet$ be a prediction function that predicts $Y_{i+1}$ from $(Y_{i-m}^{i-1},X_{i-m}^{i-1})$.
To measure the quality of prediction we use the $\ell_2$-loss (or distance), e.g., mean-square-error. 
Thus, the model order is estimated via:
\begin{align}
	\hat{m} = \argmin_{m \in \mathcal{M}} \E \left\{ d_2(Y_{i+1}, \varphi((Y_{i-m}^{i-1},X_{i-m}^{i-1})) \right\}, \label{eq:markovOrdEst}
\end{align}

\noindent where the expectation averages over everything that is random, and can be approximated via averaging.
Note that any prediction method can be used in \eqref{eq:markovOrdEst}. In fact, in \cite{Murin17BME} it was proposed to use an ensemble of predictors in order to increase the prediction power. Yet, this comes at the cost of much higher computational complexity. 
Moreover, the numerical study in \cite{Murin17BME} indicates that when the Markov order $m$ is small, $k$-NN is a very efficient predictor that out-performs significantly more complicated models such as support-vector-regression and regression based on non-linear terms. 
On the other hand, when $m$ is large (and $N$ is fixed), $k$-NN suffers from the {\em curse of dimensionality} \cite[Section 6.3]{Hastie09} and performs poorly. Note that, as stated in \cite[Sec. V.E]{Jiao15}, the number of samples required for accurate estimation of DI grows {\em exponentially} with the Markov order $m$ (this follows as increasing $m$ can be viewed as increasing the state space). Therefore, as in many  practical settings the number of samples is limited \cite{Malladi2016, MurinPCB, asilomar16}, we focus on settings where $m$ is relatively small.\footnote{When this does not hold one can either down-sample the observed sequences or decimate $(Y_{i-m}^{i-1},X_{i-m}^{i-1})$ when estimating the DI, see the discussion in \cite{Malladi2016}}. 

\begin{figure}[t]
	\captionsetup{font=footnotesize}
	\begin{center}
		\includegraphics[width=0.95\textwidth,keepaspectratio]{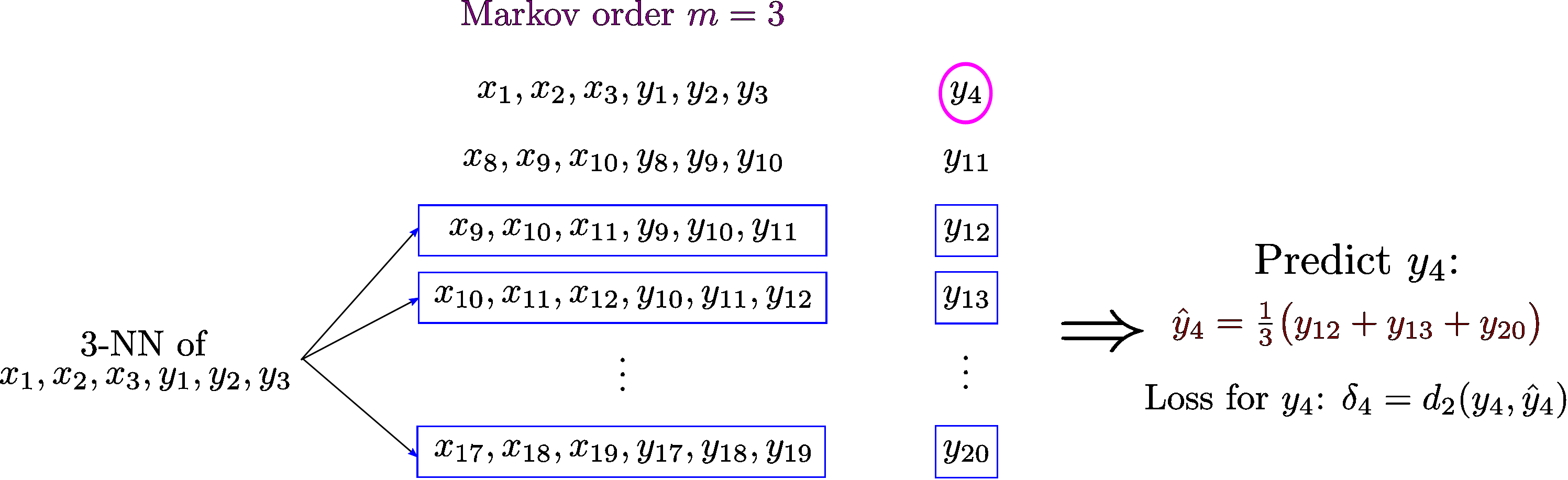}
	\end{center}
	\caption{\label{fig:NNpred} \textbf{3-NN prediction}. Illustration of 3-NN prediction of $y_4$, assuming $m=3$. }
\end{figure}

Fig. \ref{fig:NNpred} illustrates the procedure for estimating $y_4$, for $k=3$ and $m=3$. The estimator first finds the $k=3$ tuples that are nearest to the tuple $(x_1,x_2,x_3,y_1,y_2,y_3)$. Each of these tuples has a response variable; $y_{12}, y_{13}$ and $y_{20}$ in Fig. \ref{fig:NNpred}. These responses are used to predict $y_4$, resulting in the loss $\delta_4$. Note that instead of averaging  one can use weighted averaging based on the distances from $(x_1,x_2,x_3,y_1,y_2,y_3)$.
This procedure is repeated for every tuple (note that we do not search NN among the tuples that overlap with the current tuple). The calculated loss values $\delta_i$ are averaged resulting in $\Delta_m, m\in\mathcal{M}$. This procedure is repeated for all $m \in \mathcal{M}$, and the value that minimize the average loss is declared as $\hat{m}$. 

Next, we discuss the estimation of the DI measure, while assuming that the Markov order $m$ was correctly estimated.

\subsection{Estimating Directed Information}

To simplify the notation we let $X_i^- \triangleq X_{i-m}^{i-1}$ denote the past of $X_i$, and $Y_i^- \triangleq Y_{i-m}^{i-1}$ denote the past of $Y_i$. Using this notation, we write \eqref{eq:DIrateContMarkov} as:
\begin{align}
	I(\Xvec \to \Yvec) & = I(X_{i-m}^{i-1}; Y_{i}|Y_{i-m}^{i-1}) \nonumber \\
	& = h(Y, Y^{-}) - h(Y^{-}) - h(Y, Y^{-}, X^{-}) + h(Y^{-}, X^{-}). \label{eq:DIentTerms}
\end{align}

\subsubsection{The KSG Estimator}
We begin with an estimator that extends the approach used to estimate MI in \eqref{eq:MIKSG}. This estimator was presented in \cite{wibral2014} for estimating the TE measure.  
Let $\rho_{k,i,p}(X^-,Y^-,Y)$ denote the distance from {\em the tuple} $(X_i^{-}, Y_i^{-},Y_i)$ to its $k$-NN. In the following we refer to this distance as $\rho_{k,i,p}$. 
Recalling that the dimensions of $X_i^{-}$ and $Y_i^{-}$ are $m$, we note that $\rho_{k,i,p}$ is calculated in a space with dimension $2m+1$. Similarly to \eqref{eq:EntKSG}, the individual entropies are estimated using $\rho_{k,i,p}$, the distance calculated in the largest space $(X_i^{-}, Y_i^{-},Y_i)$. The resulting estimators of the individual entropy terms are then given by:
\begin{subequations} \label{eq:DIentTermsKSG}
\begin{align} 
	\hat{h}_{\text{KSG}}(Y^{-}) & = \frac{1}{N-m} \sum_{i= m+1}^N \big( \log N + \log c_{m,\infty} + m \log \rho_{k,i,\infty} - \psi(n_{Y^-, i, \infty} + 1) \big) \label{eq:DIentTermsKSG_1} \\
	\hat{h}_{\text{KSG}}(Y^{-}, Y) & = \frac{1}{N-m} \sum_{i= m+1}^N \big(\log N + \log ( c_{m,\infty} \cdot c_{1,\infty}) \nonumber \\
	 & \mspace{160mu} + (m+1) \log \rho_{k,i,\infty} - \psi(n_{(Y^-,Y), i, \infty} + 1) \big) \label{eq:DIentTermsKSG_2}
	\end{align}
	\begin{align}
	\hat{h}_{\text{KSG}}(Y^{-}, X^{-}) & = \frac{1}{N-m} \sum_{i= m+1}^N \big(\log N + \log ( c_{m,\infty} \cdot c_{m,\infty}) \nonumber \\
	 & \mspace{160mu} + 2m \log \rho_{k,i,\infty} - \psi(n_{(Y^-,X^-), i, \infty} + 1) \big) \label{eq:DIentTermsKSG_3} \\
	\hat{h}_{\text{KSG}}(Y^{-}, X^{-}, Y) & = \frac{1}{N-m} \sum_{i= m+1}^N \big(\log N + \log ( c_{m,\infty} \cdot c_{m,\infty} \cdot c_{1,\infty}) \nonumber \\
	 & \mspace{160mu} + (2m+1) \log \rho_{k,i,\infty} - \psi(k)\big). \label{eq:DIentTermsKSG_jnt}
\end{align}
\end{subequations}

\noindent Combining the entropy estimators in \eqref{eq:DIentTermsKSG} we obtain:
\begin{align}
	\mspace{-10mu} \hat{I}_{\text{KSG}}(\Xvec \mspace{-3mu} \to \mspace{-3mu} \Yvec) \mspace{-3mu} = \mspace{-3mu} \psi(k) \mspace{-3mu} + \mspace{-3mu}  \frac{1}{N \mspace{-3mu} - \mspace{-3mu} m} \mspace{-3mu} \sum_{i= m+1}^N \mspace{-8mu} \big( \psi(n_{Y^-, i, \infty} \mspace{-3mu} + \mspace{-3mu} 1) \mspace{-3mu} - \mspace{-3mu} \psi(n_{(Y^-,Y), i, \infty} \mspace{-3mu} + \mspace{-3mu} 1) \mspace{-3mu} - \mspace{-3mu} \psi(n_{(Y^-,X^-), i, \infty} \mspace{-3mu} + \mspace{-3mu} 1) \big). \label{eq:DIKSG}
\end{align}

\noindent Similarly to the KSG estimator for MI in \eqref{eq:MIKSG}, the joint entropy $h(Y^{-}, X^{-},Y)$ is estimated using the KL estimator, see \eqref{eq:DIentTermsKSG_jnt}, while the other entropy terms are estimated using sample dependent expressions, see \eqref{eq:DIentTermsKSG_1}--\eqref{eq:DIentTermsKSG_3}. 

\subsubsection{The GOV Estimator}
The second DI estimator extends the approach used to estimate MI in \eqref{eq:MIGOV}. Following the steps leading to \eqref{eq:KLEst} and \eqref{eq:EntGOV}, we obtain the following estimators:
\begin{subequations} \label{eq:DIentTermsGOV}
\begin{align} 
	\hat{h}_{\text{GOV}}(Y^{-}) & \mspace{-3mu} = \mspace{-3mu} \frac{1}{N \mspace{-3mu} - \mspace{-3mu} m} \sum_{i= m+1}^N \big( \log N \mspace{-3mu} + \mspace{-3mu} \log c_{m,2}\mspace{-3mu} + \mspace{-3mu} m \log \rho_{k,i,2} \mspace{-3mu} - \mspace{-3mu} \log(n_{Y^-, i, 2}) \big) \label{eq:DIentTermsGOV_1} \\
	\hat{h}_{\text{GOV}}(Y^{-}, Y) & \mspace{-3mu} = \mspace{-3mu} \frac{1}{N \mspace{-3mu} - \mspace{-3mu} m} \sum_{i= m+1}^N \big(\log N \mspace{-3mu} + \mspace{-3mu} \log ( c_{m+1,2}) \mspace{-3mu} + \mspace{-3mu} (m \mspace{-3mu} + \mspace{-3mu} 1) \log \rho_{k,i,2} \mspace{-3mu} - \mspace{-3mu} \log(n_{(Y^-,Y), i, 2}) \big) \label{eq:DIentTermsGOV_2} 
\end{align}
\begin{align}	
	\hat{h}_{\text{GOV}}(Y^{-}, X^{-}) & \mspace{-3mu} = \mspace{-3mu} \frac{1}{N \mspace{-3mu} - \mspace{-3mu} m} \sum_{i= m+1}^N \big(\log N \mspace{-3mu} + \mspace{-3mu} \log ( c_{2m,2}) \mspace{-3mu} + \mspace{-3mu} 2m \log \rho_{k,i,2} \mspace{-3mu} - \mspace{-3mu} \log(n_{(Y^-,X^-), i, 2}) \big) \label{eq:DIentTermsGOV_3} \\
	\hat{h}_{\text{GOV}}(Y^{-}, X^{-}, Y) & \mspace{-3mu} = \mspace{-3mu} \frac{1}{N \mspace{-3mu} - \mspace{-3mu} m} \sum_{i= m+1}^N \big(\log N \mspace{-3mu} + \mspace{-3mu} \log ( c_{2m+1,2}) \mspace{-3mu} + \mspace{-3mu} (2m \mspace{-3mu} + \mspace{-3mu} 1) \log \rho_{k,i,2} \mspace{-3mu} - \mspace{-3mu} \psi(k)\big). \label{eq:DIentTermsGOV_jnt}
\end{align}
\end{subequations}

\noindent Combining the entropy estimators in \eqref{eq:DIentTermsGOV} we obtain:
\begin{align}
	\mspace{-10mu} \hat{I}_{\text{GOV}}(\Xvec \mspace{-3mu} \to \mspace{-3mu} \Yvec) \mspace{-3mu} & = \mspace{-3mu} \psi(k) \mspace{-3mu} + \mspace{-3mu} \log \frac{c_{m+1,2} \cdot c_{2m,2}}{c_{2m+1,2} \cdot c_{1,2}} \mspace{-3mu} \nonumber \\
	& \mspace{50mu} + \mspace{-3mu}  \frac{1}{N \mspace{-3mu} - \mspace{-3mu} m} \mspace{-3mu} \sum_{i= m+1}^N \mspace{-8mu} \big( \log(n_{Y^-, i, 2}) \mspace{-3mu} - \mspace{-3mu} \log(n_{(Y^-,Y), i, 2}) \mspace{-3mu} - \mspace{-3mu} \log(n_{(Y^-,X^-), i, 2} ) \big). \label{eq:DIGOV}
\end{align}

\subsubsection{On the Statistical Significance of the Estimated DI} \label{subsubsec:sig}

Since $I(\vec{X} \to \vec{Y})$ is {\em estimated} from a {\em finite} number of samples, one may want to assess the statistical significance of this estimation. A lack of statistical significance may imply that there is no causal influence between the underlying random processes, and the estimated values are due to either noise or estimation error. 
In such a case one may choose to set the estimated DI to zero.
While for estimating GC there are known methods for quantifying the statistical significance via the respective null-distribution \cite[Sec. 2.5]{Barnett14}, the null-distribution is not known for non-parametric estimation of DI from continuous alphabet sequences.
An alternative method for evaluating the statistical significance is via a non-parametric bootstrapping procedure in the spirit of \cite{Diks2001}. 
Specifically, the idea is to {\em randomly} shuffle and re-sample $X_1^N$ such that the causal interactions from $\vec{X}$ to $\vec{Y}$ are destroyed ($\vec{Y}$ is not changed). 
We repeat this shuffle and re-sample procedure $L$ times and for each $l \in \{1,2,\dots,L\}$ we estimate the DI $I(\vec{X}(l) \to \vec{Y})$, where $\vec{X}(l)$ denotes the shuffled sequence. Since this construction destroys the causal influence, the $L$ new estimated DI values are assumed to be taken from the null-distribution of no causal influence. 
The statistical significance is then determined by the resulting P-value with parameter $\varepsilon_p$.
Finally, we note that the main drawback of such a bootstrapping procedure is the significant increase in computational complexity, since applying such a procedure amounts to multiplying the computational complexity by a factor of at least 20 (for the common value of $\varepsilon_p = 0.05$).

\subsubsection{The JVHW Estimator}

The above KSG and GOV estimators allow for a continuous input alphabet. A different approach is to first quantize the observed sequences into only a few bins in order to estimate an empirical probability mass functions (PMFs), and then estimate the DI from these PMFs. 
Such an estimator was developed by Jiao, Venkat, Han, and Weissman (JVHW) \cite{JVHW-code}, based on min-max optimal estimation of the entropy of discrete distributions \cite{Jiao15}. Specifically, the estimator uses the relationship \eqref{eq:DIentTerms} to independently estimate each of the entropy terms. 
Its main advantages are the proven optimality in estimating the entropy terms (of the discrete distributions) as well as its linear time complexity (in contrast to the universal estimators detailed in \cite{Jiao13}). 
On the other hand, when the number of available samples is small, quantizing the input sequences can lead to significant performance degradation (see the numerical study in Section \ref{sec:numerical}). 
To quantize the observed sequences one may use the optimal Lloyd-Max scalar quantizer. Our numerical study showed that when the cardinality of the discrete alphabet is large enough, Lloyd-Max quantization yields performance very similar to a naive quantization based on equal-size bins.\footnote{Here we assumed that the probability of observing samples with very large magnitude is negligible.}

To evaluate the statistical significance of the estimated values we use the same bootstrapping procedure discussed in the previous section. Note that while for discrete-alphabet samples the work \cite{yiannis16} derived the asymptotic ($N \to \infty$) null-distribution, this result may not be used when the number of samples is small.

\section{Numerical Study} \label{sec:numerical}

In this section we examine the performance of the discussed estimators in several simulated scenarios. We also consider an estimation of the GC via the toolbox \cite{Barnett14}. This is motivated by the fact that GC is widely used to quantify causal influence, and in particular when $\vec{X}$ and $\vec{Y}$ are generated through a multivariate auto-regressive process, then GC is equal to twice the DI \cite{nimaThesis}.
We consider linear as well as non-linear scenarios, where some of the scenarios were taken from \cite{Malladi2016}, and focus on the regime where $N$ is at the order of few thousands (in contrast to \cite{Malladi2016} that focused on the case of $N = 10^5$). Relatively small values of $N$ are motivated by practical applications where the number of samples is limited, e.g., \cite{MurinPCB, Theo18}.
Our simulation results indicate that the discussed $k$-NN estimators (KSG and GOV) achieve roughly the same accuracy as the estimator presented in \cite{Malladi2016}, but with an order of magnitude less samples. 

\subsection{Linear Interaction} \label{subsec:LinModel}

Let $x_i, z_i, i=1,2,\dots,N$, be i.i.d. zero-mean and unit-variance Gaussian RVs, where $x_i$ is independent of $z_i$. We generate the sequence $\{y_i\}_{i=1}^N$ via:
\begin{align}
	y_i = \beta_1 x_{i-1} + \beta_2 x_{i-2} + z_i, \quad 0 \le \beta \le 1. \label{eq:linModel}
\end{align}

\noindent Observe that in \eqref{eq:linModel} $\vec{X}$ causally influence $\vec{Y}$ while $\vec{Y}$ has no causal influence on $\vec{X}$. As was shown in \cite[Appendix B]{Malladi2016}, when $\beta_1, \beta_2 \neq 0$ then $I(\vec{X} \to \vec{Y})$ is given by:
\begin{align}
	I(\vec{X} \to \vec{Y}) = \frac{1}{2} \log_2(e) \left( \log \left(|\beta_1 \beta_2|\right) + \cosh^{-1} \left(\frac{\beta_1^2 + \beta_2^2 + 1}{2 |\beta_1 \beta_2|} \right) \right). \label{eq:LinGauss}
\end{align}

\noindent When $\beta_1=0, \beta_2 \neq 0$, then  $I(\vec{X} \to \vec{Y}) = \frac{1}{2} \log_2 \left(1 + \beta_2^2 \right)$, and when $\beta_1 \neq 0, \beta_2 = 0$, then  $I(\vec{X} \to \vec{Y}) = \frac{1}{2} \log_2 \left(1 + \beta_1^2 \right)$. Here, the DI is measured in bits.

\begin{figure}[t]
    \centering
		\begin{subfigure}[t]{0.5\textwidth}
        \centering
				\captionsetup{font=footnotesize}
        \includegraphics[width=1\columnwidth,keepaspectratio]{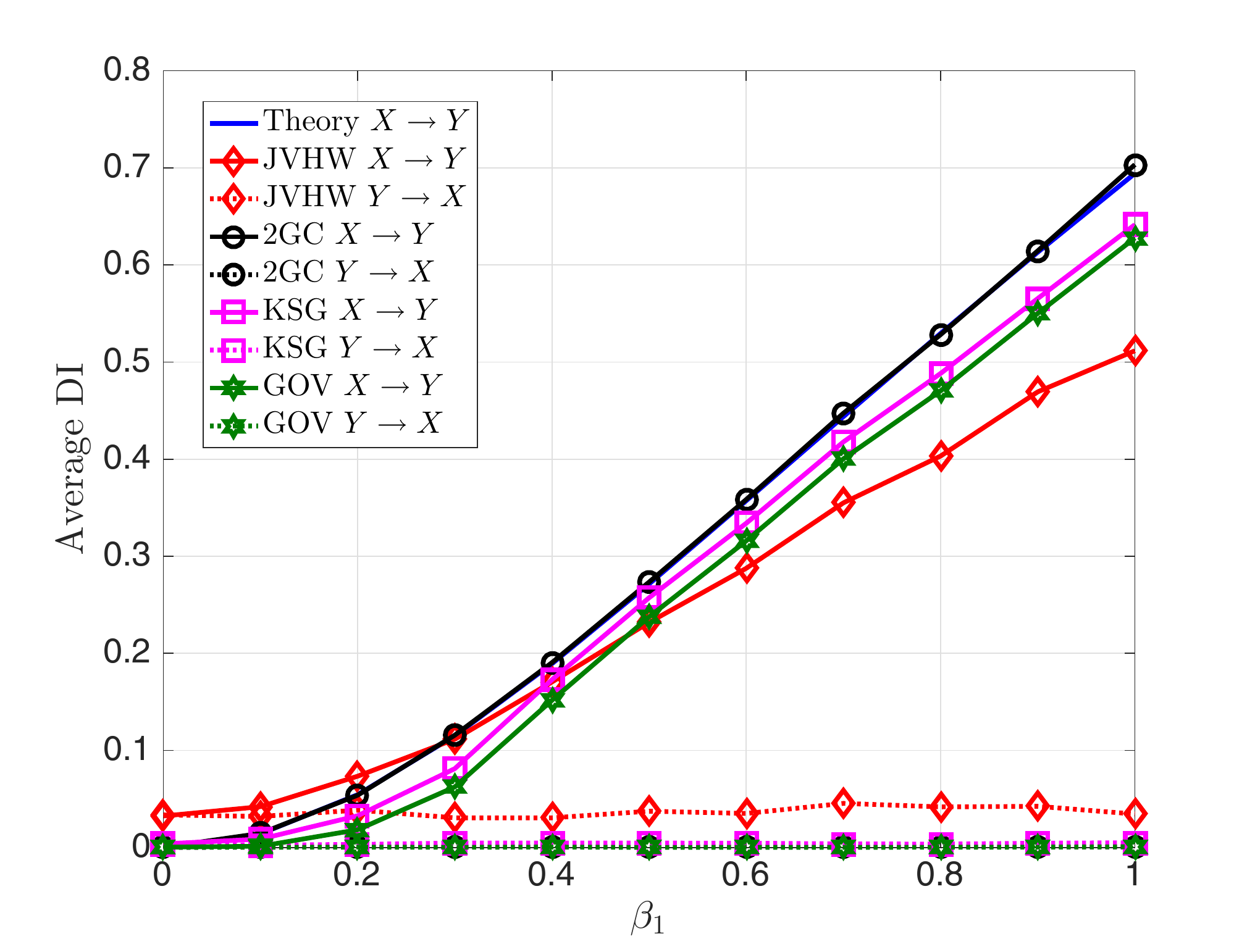}
        \vspace{-0.1cm}
				\caption{}
    \end{subfigure}%
    \begin{subfigure}[t]{0.5\textwidth}
        \centering
				\captionsetup{font=footnotesize}
        \includegraphics[width=1\columnwidth,keepaspectratio]{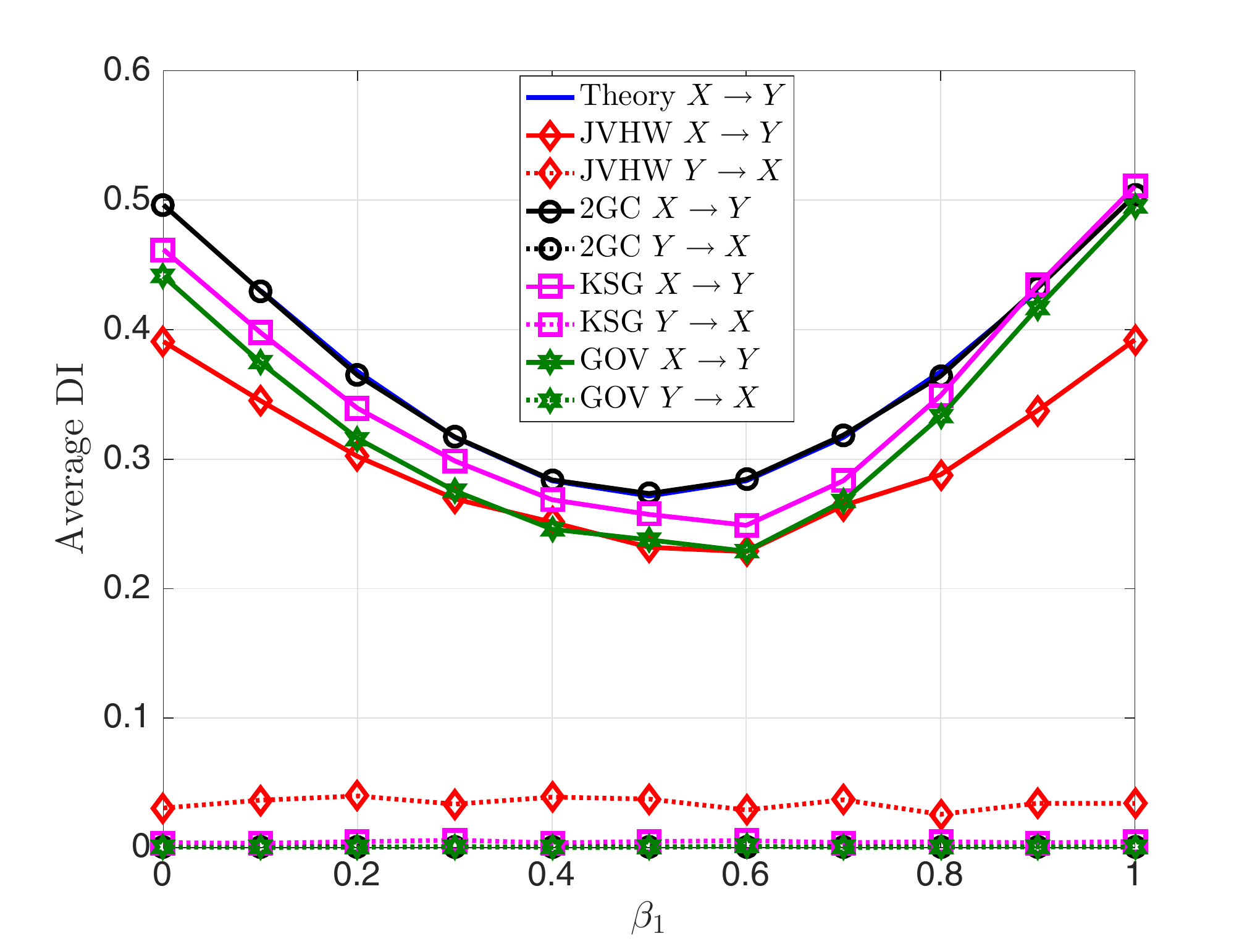}
        \vspace{-0.1cm}
				\caption{\label{fig:LinGaussEstShort_b}}
    \end{subfigure}
		\captionsetup{font=footnotesize}
		\vspace{-0.1cm}
    \caption{{\bf Average DI estimates versus $\beta_1$ for the interaction \eqref{eq:LinGauss}, $N = 3000$.} {\em Four} quantization levels are used in the JVHW estimator. In the KSG and GOV estimators $K=8$. In sub-figure (a) $\beta_2 = \beta_1$ while in sub-figure (b) $\beta_2 = 1 - \beta_1$. \label{fig:LinGaussEstShort}}
\end{figure}

Fig. \ref{fig:LinGaussEstShort} depicts the mean of the estimated DIs versus the value of $\beta_1$, for $\beta_2 = \beta_1$ and for $\beta_2 = 1-\beta_1$, where $\beta_1 \in \{0,0.1,\dots,1\}$. 
The DI values were estimated from $N = 3000$ samples, and the mean was calculated over 48 independent (with different random generator seeds) trials. For both plots in Fig. \ref{fig:LinGaussEstShort}, the average standard deviation for the GC estimator is about 0.01 for $I(\vec{X} \to \vec{Y})$ and 0.0006 for $I(\vec{Y} \to \vec{X})$; the average standard deviation for the JVHW estimator is about 0.05 for $I(\vec{X} \to \vec{Y})$ and 0.03 for $I(\vec{Y} \to \vec{X})$; the average standard deviation for the KSG estimator is about 0.02 for $I(\vec{X} \to \vec{Y})$ and 0.006 for $I(\vec{Y} \to \vec{X})$; and, the average standard deviation for the GOV estimator is about 0.02 for $I(\vec{X} \to \vec{Y})$ and 0.002 for $I(\vec{Y} \to \vec{X})$. 
The statistical significance of the estimated values was evaluated using the approach discussed in Section \ref{subsubsec:sig} (for the GC estimates we used the null-distribution as stated in \cite[Sec. 2.5]{Barnett14}). 
For $\beta_2 = \beta_1$ and $I(\vec{X} \to \vec{Y})$, when $\beta_1 \ge 0.5$, all the estimators yield significant estimations in all trials. While the GC estimator yield significant estimations in all trials when $\beta_1 \ge 0.1$, the KSG and GOV estimators yield significant estimations in all trials when $\beta_1 \ge 0.3$. For $\beta_1 = 0.1$ and $\beta_1 = 0.2$ the KSG and GOV estimators yield significant estimations in about 20\% and 90\% of the trials, respectively. The significance of the JVHW estimator gradually increases to 100\% in the range $0 \le \beta_1 \le 0.5$. 
For $\beta_2 = \beta_1$ and $I(\vec{X} \to \vec{Y})$, as the true DI values are relatively large, all the estimators yield significant estimations in all the trials.

Comparing the estimation results in Fig. \ref{fig:LinGaussEstShort} to the estimation results reported in \cite[Fig. 2]{Malladi2016}, one can observe that the curves are very similar (KSG and GOV versus the KDE estimator of \cite{Malladi2016}). 
Yet, there is a fundamental difference between the estimators: in \cite[Fig. 2]{Malladi2016} the number of samples is $N=10^5$, while in the above Fig. \ref{fig:LinGaussEstShort} the number of samples is $N=3000$. Thus, in the considered scenario, to achieve the same accuracy {\em the KSG and GOV estimators require an order of magnitude less samples compared to the KDE estimator}. 

\begin{figure}[t]
    \centering
		\begin{subfigure}[t]{0.5\textwidth}
        \centering
				\captionsetup{font=footnotesize}
        \includegraphics[width=1\columnwidth,keepaspectratio]{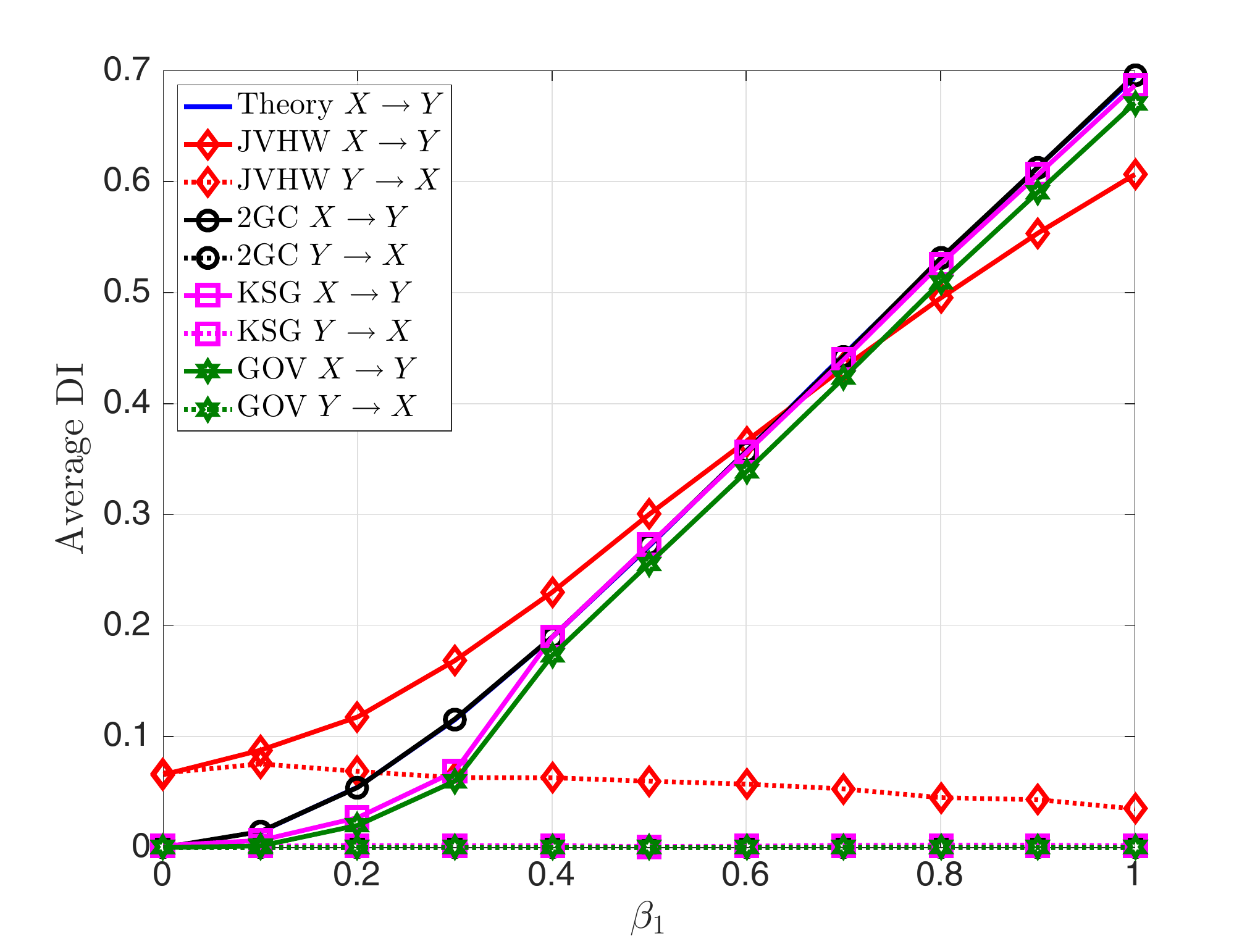}
        \vspace{-0.1cm}
				\caption{\label{subfig:LinGaussLongAll}}
    \end{subfigure}%
    \begin{subfigure}[t]{0.5\textwidth}
        \centering
				\captionsetup{font=footnotesize}
        \includegraphics[width=1\columnwidth,keepaspectratio]{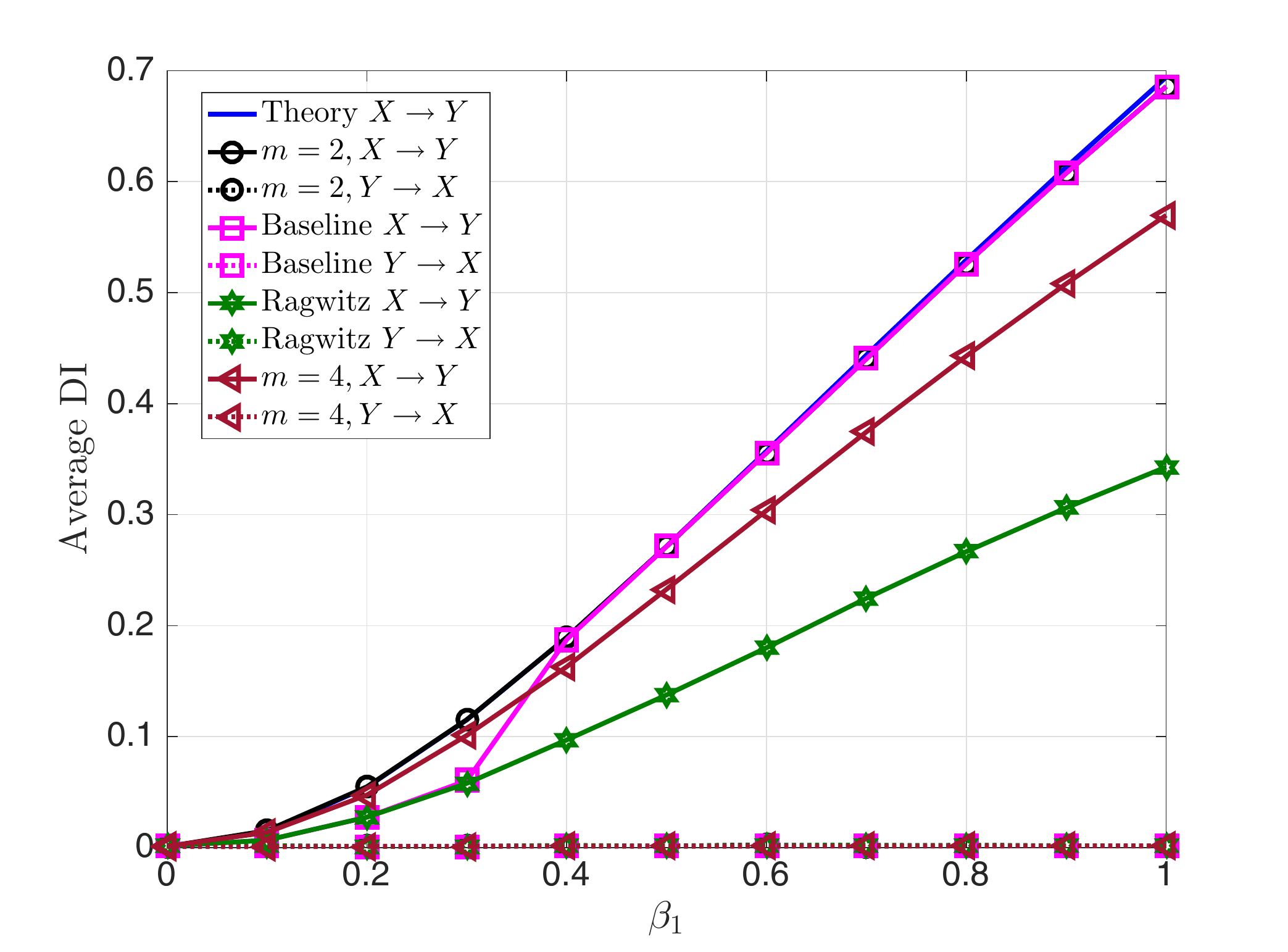}
        \vspace{-0.1cm}
				\caption{\label{subfig:LinGaussLongModOrd}}
    \end{subfigure}
		\captionsetup{font=footnotesize}
		\vspace{-0.1cm}
    \caption{{\bf Average DI estimates versus $\beta_1$ for the interaction \eqref{eq:linModel}, $N = 20000$.} {\em Six} quantization levels are used in the JVHW estimator. In the KSG and GOV estimators $K=8$. 
		(a) Average estimated DI of the JVHW, GC, KSG, and GOV estimators, when $\beta_2 = \beta_1$. (b) Average estimated DI, of the KSG estimator with different Markov order estimation methods, when $\beta_2 = \beta_1$
		\label{fig:LinGaussEstLong}}
\end{figure}

To explore the performance of the discussed estimators when $N$ is larger, we repeated the experiment specified in \eqref{eq:linModel} when $N = 20000$, and $\beta_2 = \beta_1$. The results are depicted in Fig. \ref{fig:LinGaussEstLong}. Observing Fig. \ref{subfig:LinGaussLongAll} we note that the accuracy of the KSG and GOV estimator is significantly higher (compared to Fig. \ref{fig:LinGaussEstShort}) when $\beta_1 \ge 0.4$. Moreover, the standard deviations of the estimated values are smaller by about an order of magnitude. On the other hand, for $0.1 \le \beta_1 \le 0.3$, one can observe an apparent bias. 
This bias is due to an inaccurate estimation of the Markov order. When $\beta_1 \le 0.3$ the noise level is significantly higher than the signal level and the estimation method discussed in Section \ref{subsec:MarkovOrdEst} under estimates the true Markov order. To verify this observation, in Fig. \ref{subfig:LinGaussLongModOrd} we present the average estimated DI, for the KSG estimator, and for several alternatives for the Markov order estimation:
\begin{itemize}
	\item 
		The approach discussed in Section \ref{subsec:MarkovOrdEst}. 
		The corresponding curve is denoted by ``Baseline''. 
		
	\item
		The approach proposed in \cite{Ragwitz2002} where a $k$-NN estimator is used to estimate the Markov order based {\em only} on the past samples of $\vec{Y}$, ignoring the past samples of $\vec{X}$.
		The corresponding curve is denoted by ``Ragwitz''. 
		
	\item
		Fixing the Markov order to the correct order according to \eqref{eq:linModel}, i.e., setting $m=2$.
		The corresponding curve is denoted by ``$m=2$''. 
		
	\item
		Setting the Markov order to $m=4$. 
		The corresponding curve is denoted by ``$m=4$''. 
\end{itemize}

It can be observed that the method by \cite{Ragwitz2002} performs much worse than the method discussed in Section \ref{subsec:MarkovOrdEst}. In fact, the curve corresponding to the ``Ragwitz'' method is almost identical to a curve generated by setting $m=1$. Setting $m=2$, the true order, leads to almost no errors comparing to the theoretical values. Finally, setting $m=4$ works better for small values of $\beta_1$ (when the problem of estimating the Markov order is very challenging), yet a significant performance degradation can be observed for larger values of $\beta_1$. 

\subsection{Quadratic Interaction}

Next, we consider a quadratic dependency between $\vec{X}$ and $\vec{Y}$. 
Again, we let $x_i, z_i, i=1,2,\dots,N$ be i.i.d. zero-mean and unit-variance Gaussian RVs, where $x_i$ is independent of $z_i$. We generate the sequence $\{y_i\}_{i=1}^N$ via:
\begin{align}
	y_i = \beta_1 x_{i-1}^2 + \beta_2 x_{i-2}^2 + z_i. \label{eq:quadModel}
\end{align}

\noindent A similar model was studied in \cite[Sec. V.B]{Malladi2016}. Similarly to Section \ref{subsec:LinModel}, we consider two scenarios: $\beta_2 = \beta_1$ and $\beta_2 = 1 - \beta_1$. When $\beta_2 = \beta_1$ we expect $I(\vec{X} \to \vec{Y})$ to increase monotonically with $\beta_1$, and when $\beta_2 = 1 - \beta_1$ we expect to observe a ``U'' shape as in Fig. \ref{fig:LinGaussEstShort_b}. Similarly to Fig. \ref{fig:LinGaussEstShort} there should be no causal influence from $\vec{Y}$ to $\vec{X}$. We note that calculating the theoretical DI values of this model seems intractable. 

\begin{figure}[t]
    \centering
		\begin{subfigure}[t]{0.5\textwidth}
        \centering
				\captionsetup{font=footnotesize}
        \includegraphics[width=1\columnwidth,keepaspectratio]{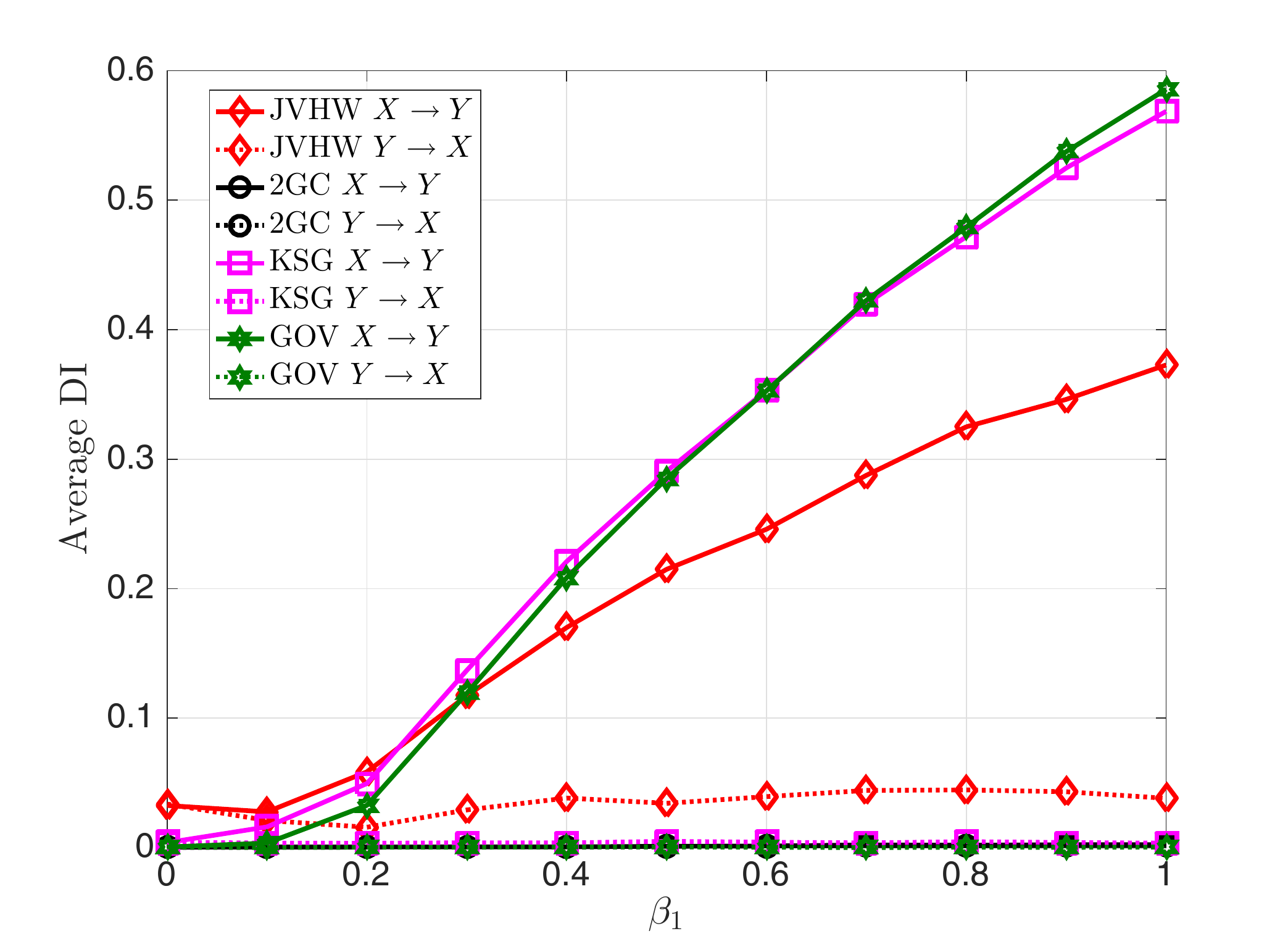}
        \vspace{-0.1cm}
				\caption{}
    \end{subfigure}%
    \begin{subfigure}[t]{0.5\textwidth}
        \centering
				\captionsetup{font=footnotesize}
        \includegraphics[width=1\columnwidth,keepaspectratio]{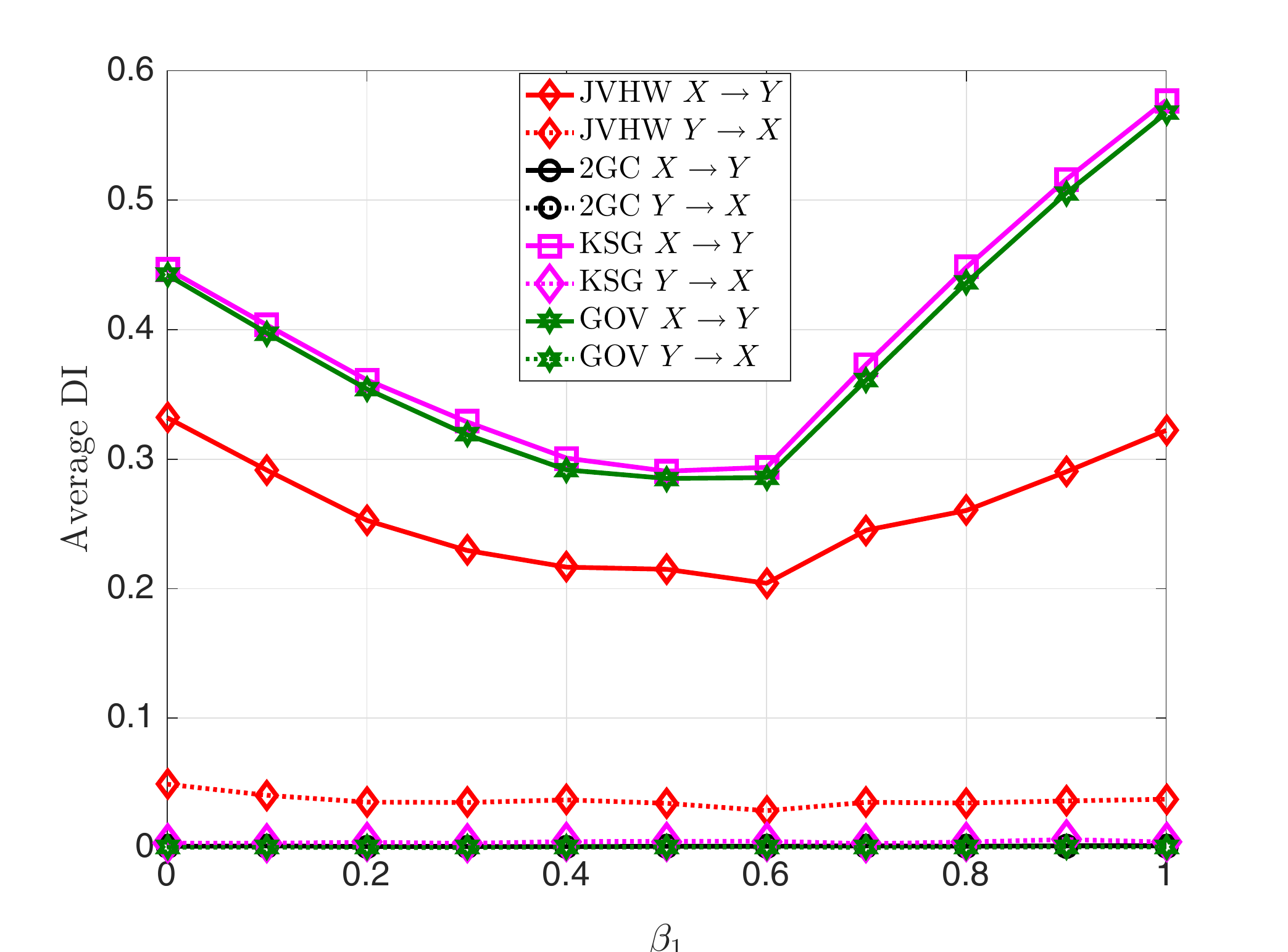}
        \vspace{-0.1cm}
				\caption{}
    \end{subfigure}
		\captionsetup{font=footnotesize}
		\vspace{-0.1cm}
    \caption{{\bf Average DI estimates versus $\beta_1$ for the interaction \eqref{eq:quadModel}, $N = 3000$.} {\em Four} quantization levels are used in the JVHW estimator. In the KSG and GOV estimators $K=8$. In sub-figure (a) $\beta_2 = \beta_1$ while in sub-figure (b) $\beta_2 = 1 - \beta_1$. \label{fig:QuadGaussEstShort}}
\end{figure}

Fig. \ref{fig:QuadGaussEstShort} depicts the mean of the estimated DIs versus the value of $\beta_1$, for $\beta_2 = \beta_1$ and for $\beta_2 = 1-\beta_1$.
$N = 3000$ samples were used to estimated the DI values. The mean values reported in Fig. \ref{fig:QuadGaussEstShort} were calculated over 48 independent trials. Similarly to the linear scenario, for both plots the average standard deviation for the GC estimator is about 0.001 for $I(\vec{X} \to \vec{Y})$ and 0.0008 for $I(\vec{Y} \to \vec{X})$; the average standard deviation for the JVHW estimator is about 0.04 for $I(\vec{X} \to \vec{Y})$ and 0.03 for $I(\vec{Y} \to \vec{X})$; the average standard deviation for the KSG estimator is about 0.02 for $I(\vec{X} \to \vec{Y})$ and 0.006 for $I(\vec{Y} \to \vec{X})$; and, the average standard deviation for the GOV estimator is about 0.02 for $I(\vec{X} \to \vec{Y})$ and 0.002 for $I(\vec{Y} \to \vec{X})$. Except the GC estimator, the statistical significance is also similar to the linear scenario. 
It can be observed that as expected when $\beta_2 = \beta_1$, the estimated DI monotonically increases with $\beta_1$. Moreover, the results are similar to \cite[Fig. 3]{Malladi2016}, again with the exception that significantly smaller number of samples was used for estimation ($N = 3000$ versus $N = 10^5$). 
Finally, Fig. \ref{fig:QuadGaussEstShort} indicates that the JVHW estimator is able to capture the causal influence from $\vec{X}$ to $\vec{Y}$, although the estimated values are smaller than those estimated by the $k$-NN estimators. On the other hand, as expected and as indicated in \cite{Malladi2016}, GC is not able to capture the causal influence from $\vec{X}$ to $\vec{Y}$. This follows as GC is based on a linear model while \eqref{eq:quadModel} obeys a quadratic one.

Next, we discuss two non-linear coupling maps that were discussed in \cite{Ishiguro2008}.

\subsection{Noisy H\'enon Map}

We first study a model that consists of unidirectional coupling via the H\'enon map \cite{Henon76}. Let $\tilde{x}_j, \tilde{y}_j \sim \mathcal{U}(0,1), j=1,2$, and let $\tilde{N} = 10^5 + 3000$. We generate the sequence $\{\tilde{y}_i, \tilde{x}_i\}_{i=3}^{\tilde{N}}$ via:
\begin{align}
	\tilde{x}_{i+2} & = 1.4 - \tilde{x}_{i+1}^2 + 0.3 \tilde{x}_i, \nonumber \\
	\tilde{y}_{i+2} & = 1.4 - (\beta \tilde{x}_{i+1} + (1-\beta)\tilde{y}_{i+1})\tilde{y}_{i+1} + 0.3 \tilde{y}_i, \label{eq:henon_basic}
\end{align}

\noindent for $0 \le \beta \le 1$. 
Next, let $z_{x,i}, z_{y,i}, i=1,2,\dots,\tilde{N}$ be i.i.d. zero-mean and unit-variance Gaussian RVs, where $z_{x,i}$ is independent of $z_{y,i}$. We generate the sequence $\{y_i, x_i\}_{i=1}^{\tilde{N}}$ via:
\begin{align}
	y_i = \tilde{y}_i +  \gamma \cdot z_{y,i}, \quad x_i = \tilde{x}_i + \gamma \cdot z_{x,i}, \label{eq:henon_noisy}
\end{align}

\noindent where $\gamma$ is a gain parameter that sets the signal-to-noise-ratio. The DI is estimated from the {\em last} $N=3000$ samples of the sequences $\{y_i, x_i\}$.

\begin{figure}[t]
    \centering
		\begin{subfigure}[t]{0.5\textwidth}
        \centering
				\captionsetup{font=footnotesize}
        \includegraphics[width=1\columnwidth,keepaspectratio]{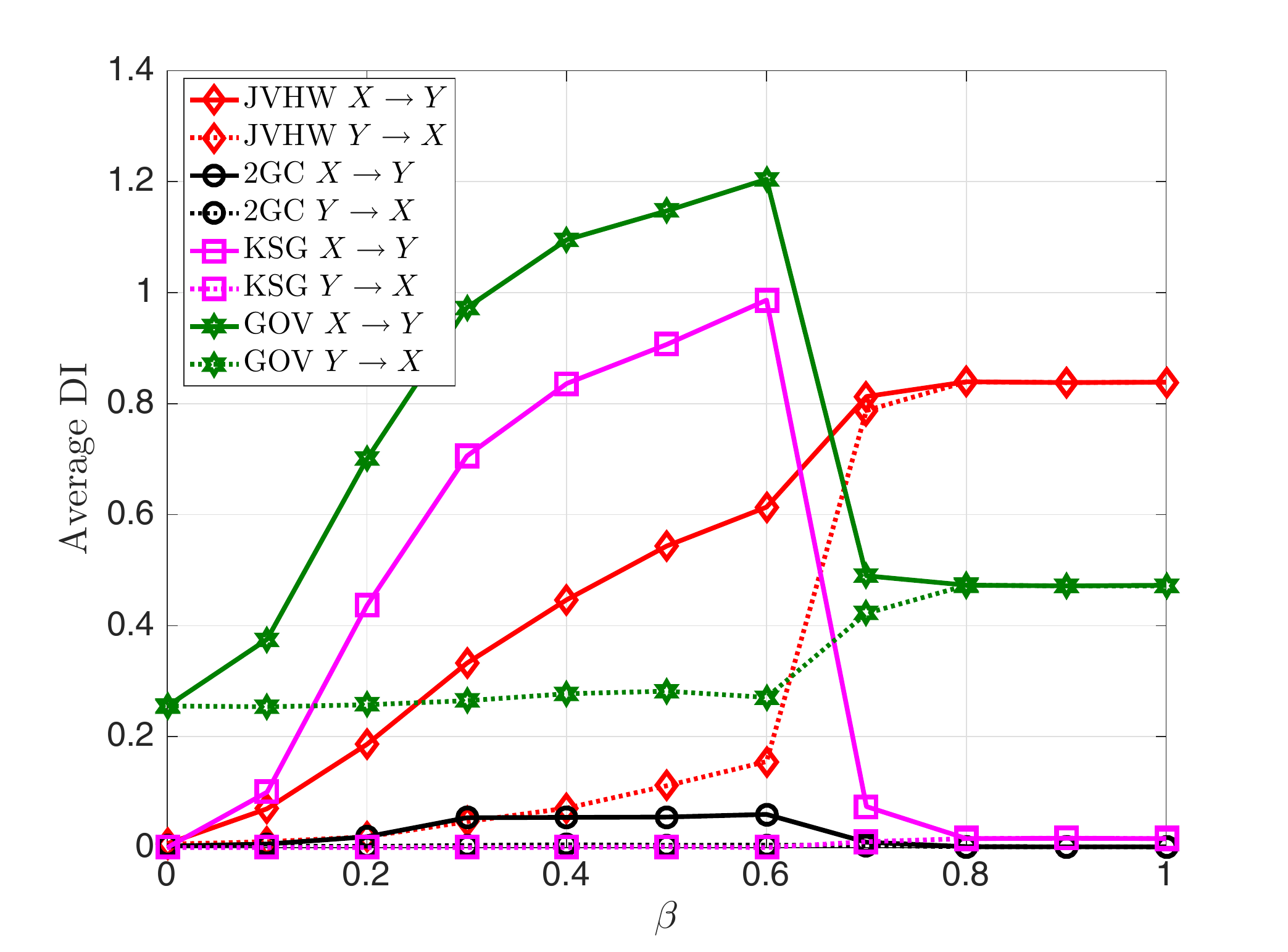}
        \vspace{-0.1cm}
				\caption{\label{subfig:henonZero}}
    \end{subfigure}%
    \begin{subfigure}[t]{0.5\textwidth}
        \centering
				\captionsetup{font=footnotesize}
        \includegraphics[width=1\columnwidth,keepaspectratio]{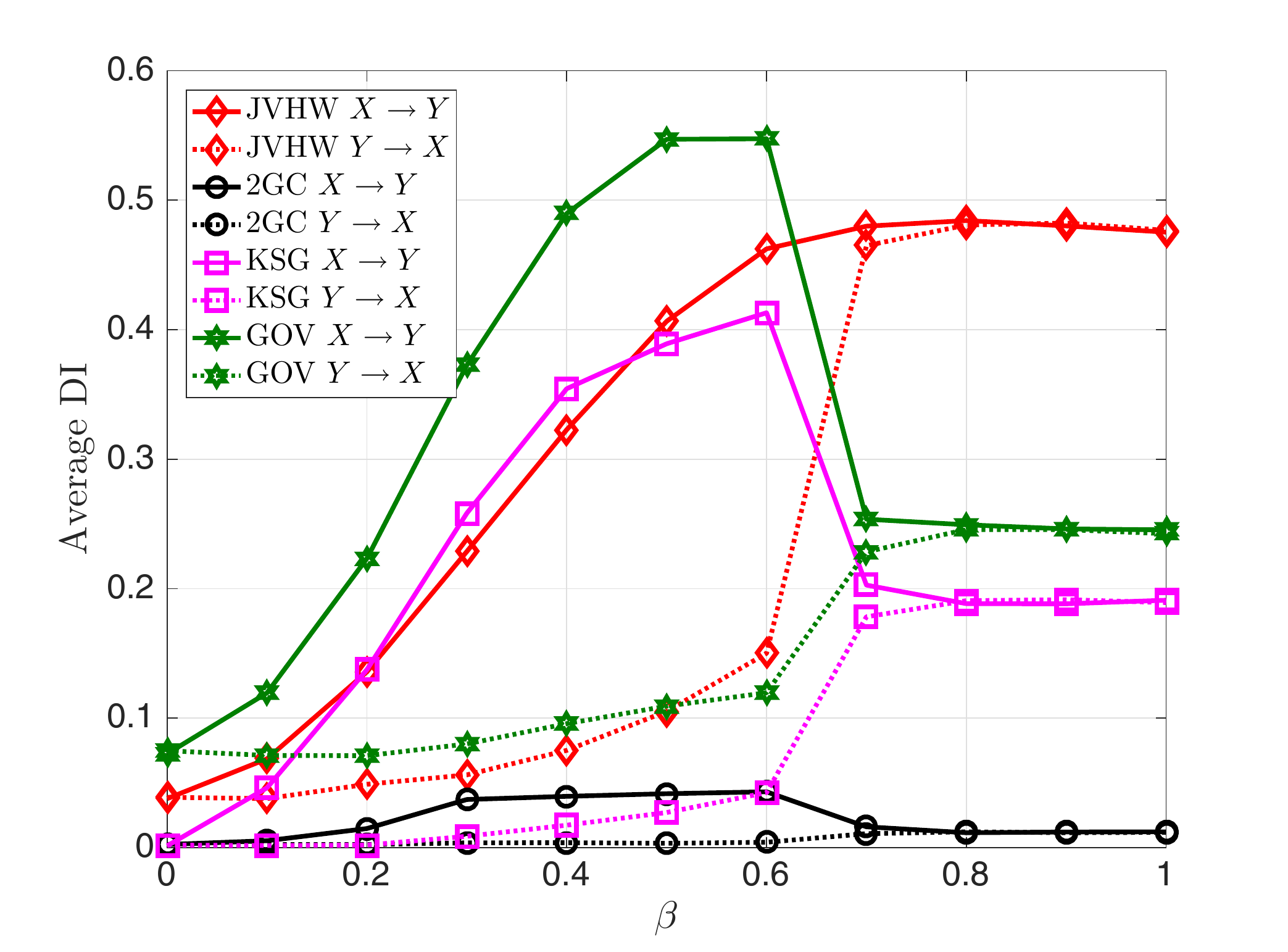}
        \vspace{-0.1cm}
				\caption{\label{subfig:henonPoint3}}
    \end{subfigure}
		\captionsetup{font=footnotesize}
		\vspace{-0.1cm}
    \caption{{\bf Average DI estimates versus $\beta$ for the interaction \eqref{eq:henon_basic}--\eqref{eq:henon_noisy}, $N = 3000$.} {\em Four} quantization levels are used in the JVHW estimator. In the KSG and GOV estimators $K=8$. 
		(a) $\gamma = 0.001$. (b) $\gamma = 0.3$.
		\label{fig:hennon}}
\end{figure}

We first note that in \eqref{eq:henon_basic} the parameter $\beta$ controls the strength of coupling between the two sequences. Moreover, \eqref{eq:henon_basic} indicates that $\vec{X}$ causally influences $\vec{Y}$, and this influence should increase with $\beta$. On the other hand, based on \eqref{eq:henon_basic} we do not expect any causal influence from $\vec{Y}$ to $\vec{X}$. 
As discussed in \cite{Ishiguro2008}, for the noiseless scenario, namely $\gamma = 0$, when $\beta > 0.7$, then $\vec{X}$ and $\vec{Y}$ are fully synchronized and the two sequences are indistinguishable. Thus, in this case the causal flow of information is zero. 
Fig. \ref{subfig:henonZero}, that considers the almost noiseless setting,\footnote{Here we do not set $\gamma$ to exactly zero in order not to violate Assumption \ref{assmp:bounded_2}, and to prevent numerical instabilities when estimating GC.} indicates that the KSG estimator indeed estimates an increasing causal influence from $\vec{X}$ to $\vec{Y}$ for $\beta < 0.7$. Moreover, this causal influence drops to almost zero when $\beta$ is increased beyond 0.7, indicating on the (almost) deterministic relationship between $\vec{X}$ and $\vec{Y}$. This result is in full correspondence to \cite[Figs. 2a and 2b]{Ishiguro2008}.
Fig. \ref{subfig:henonZero} also indicates that the KSG estimator correctly infers a negligible causal influence from $\vec{Y}$ to $\vec{X}$ for all values of $\beta$. 
It can further be observed from Fig. \ref{subfig:henonZero} that the curve corresponding to the GOV estimator is biased, and that the JVHW estimator misses the drop in causal influence for $\beta > 0.7$. 

In Fig. \ref{subfig:henonZero} we consider the case of $\gamma = 0.3$. It can be observed that for the KSG estimator the results are similar to the case of $\gamma = 0.001$, with the exception that the DI, in both directions, is significantly larger than zero for $\beta > 0$. This exactly matches the desired behavior as indicated in \cite[Figs. 4a and 4b]{Ishiguro2008}. For this setting, the GOV estimator seems to be less biased, while the JVHW still misses the decrease in causal influence for $\beta > 0.7$. 

Finally, we note that for both $\gamma = 0.001$ and $\gamma = 0.3$, GC fails to infer any significant causal influence from $\vec{X}$ to $\vec{Y}$ or from $\vec{Y}$ to $\vec{X}$.

\subsection{Sigmoid Coupling}

In this sub-section we discuss a scenario where $\vec{X}$ drives $\vec{Y}$ through a sigmoid function \cite[Sec. E]{Ishiguro2008}, defined as:
\begin{align*}
	\mathsf{Sigmoid}(\theta) = \frac{1}{1 + \exp \{-\theta\}}.
\end{align*} 

\noindent Let $x_1, y_1 \sim \mathcal{U}(0,1)$. We generate the sequence $\{y_i, x_i\}_{i=2}^{N}$ via:
\begin{align}
	x_{i+1} & = 0.125 x_i + \frac{25 x_i}{4(x_i^2 + 1)} + 2 \cos(1.2 \cdot i) + z_{x,i}, \nonumber \\
	y_{i+1} & = 0.1 y_i^2 - \beta \left( \mathsf{Sigmoid}(x_i)^2 - 0.3 \right) + z_{y,i}, \label{eq:sigmoid}
\end{align}

\noindent where $0 \le \beta \le 1$, and $z_{x,i}, z_{y,i}, i=2,3,\dots,N$, are i.i.d. zero-mean and unit-variance Gaussian RVs ($z_{x,i}$ is independent of $z_{y,i}$). \eqref{eq:sigmoid} indicates that $\vec{X}$ causally influences $\vec{Y}$, and this influence should increase with $\beta$. On the other hand, based on \eqref{eq:sigmoid} we do not expect any causal influence from $\vec{Y}$ to $\vec{X}$.

\begin{figure}[t]
    \centering
		\captionsetup{font=footnotesize}
    \includegraphics[width=0.65\columnwidth,keepaspectratio]{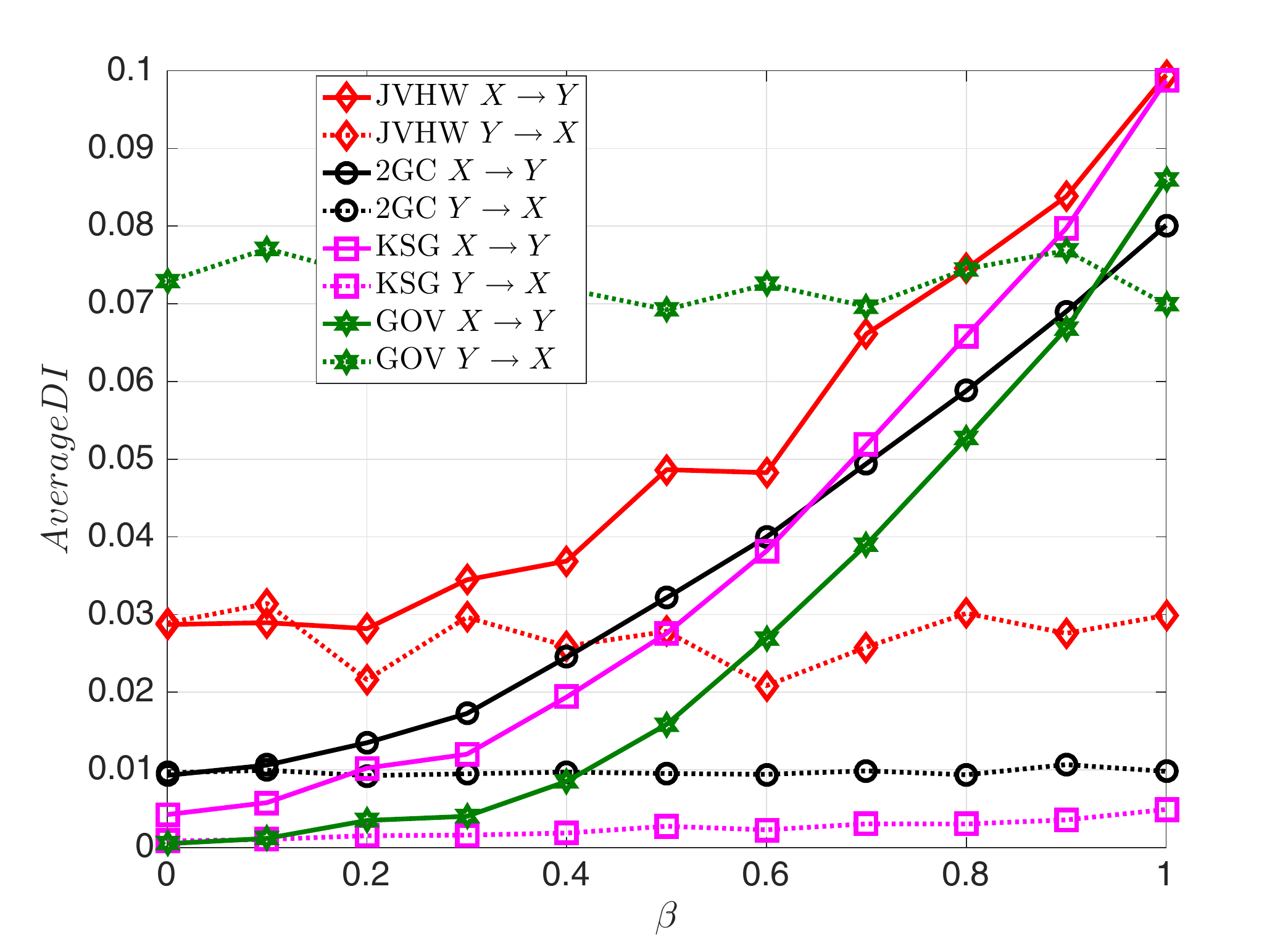}
    \vspace{-0.1cm}
		\caption{{\bf Average DI estimates versus $\beta$ for the interaction \eqref{eq:sigmoid}, $N = 3000$.} {\em Four} quantization levels are used in the JVHW estimator. In the KSG and GOV estimators $K=8$.
		\label{fig:sigmoid}}
\end{figure}

Fig. \ref{fig:sigmoid} depicts the average estimated DI versus $\beta$. It can be observed that for all four estimators (KSG, GOV, GC, and JVHW) the average $\hat{I}(\vec{X}$ to $\vec{Y})$ increases with $\beta$. We note here that the curve corresponding to the GC estimator is somewhat surprising as the interaction \eqref{eq:sigmoid} is clearly non-linear. When examining the average $\hat{I}(\vec{Y}$ to $\vec{X})$ it can be observed that both GOV and JVHW find a non-negligible causal influence (which does not exist according to \eqref{eq:sigmoid} and \cite[Sec. E]{Ishiguro2008}). On the other hand, the KSG estimator estimates a negligible amount of causal influence from $\vec{Y}$ to $\vec{X}$. Thus, the KSG estimator is the only one to follow the expected results as indicated in \cite[Figs. 11a and 11c]{Ishiguro2008}.




\end{document}